\def\year{2019}\relax
\documentclass[letterpaper]{article} 
\usepackage{aaai19}  
\usepackage{times}  
\usepackage{helvet} 
\usepackage{courier}  
\usepackage[hyphens]{url}  
\usepackage{graphicx} 
\usepackage{graphicx}  
\usepackage{mathptmx}
\usepackage{hyphenat}
\usepackage{xspace}
\usepackage{amsmath} 
\usepackage{multirow}
\usepackage{subfig}
\usepackage{makecell}
\usepackage[title]{appendix}

\newcommand{\textcite}[1]{\citeauthor{#1} \shortcite{#1}}

\newcommand{\ie}{\textit{i.e.}\xspace}
\newcommand{\eg}{\textit{e.g.}\xspace}
\newcommand{\cf}{\textit{cf.}\xspace}

\newcommand{\vs}{\textit{vs.}\xspace}
\newcommand{\etc}{\textit{etc.}\xspace}

\newcommand{\Secref}[1]{Sec.~\ref{#1}}

\newcommand{\Tabref}[1]{Table~\ref{#1}}
\newcommand{\Figref}[1]{Fig.~\ref{#1}}

\title{Eliciting New Wikipedia Users' Interests via Automatically Mined Questionnaires: 
For a Warm Welcome, Not a Cold Start}
\author{Ramtin Yazdanian,\textsuperscript{\rm 1} Leila Zia,\textsuperscript{\rm 2} Jonathan Morgan,\textsuperscript{\rm 2} Bahodir Mansurov,\textsuperscript{\rm 2} Robert West\textsuperscript{\rm 1}\\ 
\textsuperscript{\rm 1}EPFL\\
\textsuperscript{\rm 2}Wikimedia Foundation\\
ramtin.yazdanian@epf\/l.ch, leila@wikimedia.org, jmorgan@wikimedia.org, bmansurov@wikimedia.org, robert.west@epf\/l.ch 
}
 
\begin{document}

\maketitle

\begin{abstract}
Every day, thousands of users sign up as new Wikipedia contributors. Once joined, these users have to decide which articles to contribute to, which users to seek out and learn from or collaborate with, \etc Any such task is a hard and potentially frustrating one given the sheer size of Wikipedia.
Supporting newcomers in their first steps by recommending articles they would enjoy editing or editors they would enjoy collaborating with is thus a promising route toward converting them into long-term contributors.
Standard recommender systems, however, rely on users' histories of previous interactions with the platform. As such, these systems cannot make high-quality recommendations to newcomers without any previous interactions---the so-called cold-start problem.
The present paper addresses the cold-start problem on Wikipedia by developing a method for automatically building short questionnaires that, when completed by a newly registered Wikipedia user, can be used for a variety of purposes, including article recommendations that can help new editors get started.
Our questionnaires are constructed based on the text of Wikipedia articles as well as the history of contributions by the already onboarded Wikipedia editors.
We assess the quality of our questionnaire-based recommendations in an offline evaluation using historical data, as well as an online evaluation with hundreds of real Wikipedia newcomers, concluding that our method provides cohesive, human-readable questions that perform well against several baselines.
By addressing the cold-start problem, this work can help with the sustainable growth and maintenance of Wikipedia's diverse editor community. 
\end{abstract}

\section{Introduction}

Every day, roughly 10,000 users open an account on Wikipedia across its more than 160 actively edited language editions. The retention and diversification of these newcomers is crucial to Wikipedia's expansion and success \cite{doi:10.1177/0002764212469365}, as the largest encyclopedia of humankind relies on a large and diverse community of active editors.

Welcoming the newcomers to Wikipedia, however, comes with its own challenges.
One of the first challenges is to quickly learn about the newcomers' interests, a critical piece of information for building personalized recommendation systems to read and edit Wikipedia, for matching newcomers with other newcomers and with more experienced editors, for assessing the diversity of the editor pool, \etc 
But learning about newcomers' interests and profiles is hard as they do not have a history of contributions to Wikipedia or interactions with the platform yet.
This so-called \emph{cold-start problem} \cite{c0_3} is well known in the context of many recommendation algorithms such as collaborative filtering \cite{c0} or content-based filtering \cite{c0_1}, families of algorithms that rely heavily on user histories.
Generally, the cold-start problem arises when recommendations are to be made for a new user, for whom no past history exists, thus drastically reducing recommendation quality.

Many of the methods used to solve the cold-start problem are based on automatically generated questionnaires posed to the user, by means of which the system creates an initial profile for the user and then recommends items based on said profile. The main problem to be tackled, then, is how a questionnaire can be automatically generated from available data, which, in the case of Wikipedia, consists of the content of articles and the editing history of existing users, which is a form of collaborative data. In particular for Wikipedia, the challenge is that there is no explicit feedback (such as the ratings omnipresent in platforms such as Amazon \etc) available by editors on the articles they like or dislike, but only implicit feedback \cite{c1}: the only data that is publicly available about the interactions of individual users with articles is logs of the edits users have made to articles. 

In this paper, we propose a set of language\hyp independent methods for the data-driven generation of questionnaires that allow us to elicit initial profiles from Wikipedia newcomers, with an application to article recommendations. 

\begin{figure}[t]
    \resizebox{\columnwidth}{!}{
        \includegraphics{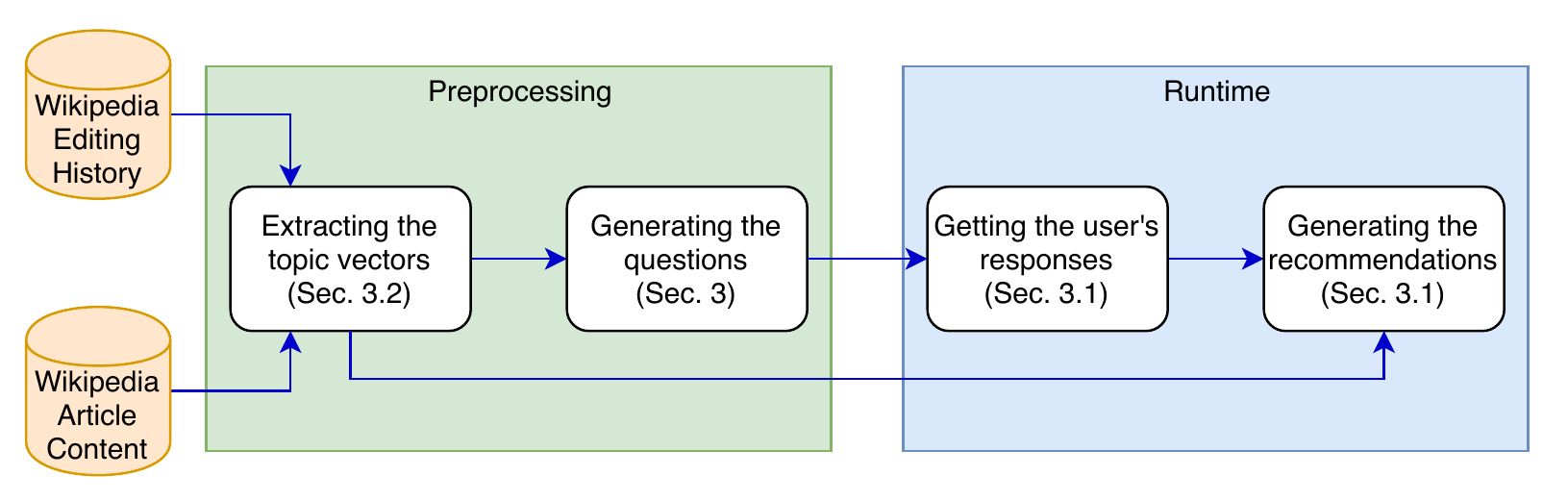}
    }
    \caption{An overview of our whole pipeline for creating the questionnaire and providing users with recommendations, along with section numbers for each element of the pipeline.}
    \label{fig:entire_pipeline}
\end{figure}

The following are our main contributions: 
(i)~We propose \textbf{a language\hyp independent question generation method} that takes topics (represented as vectors with a weight for every Wikipedia article, indicating the relevance of the respective article to the topic) as input and turns them into questions that ask the user to compare a pair of article lists and choose their preferred list.
(ii)~We propose \textbf{three topic extraction methods,} based on the content of Wikipedia articles, their editing history, or both. This provides the topics needed for our question generation method. After presenting these questions to a user and thus eliciting their initial profile, we provide them with \textbf{article recommendations} by combining their questionnaire answers with the extracted topics. Other forms of assistance (\emph{e.g.} matching newcomers to other editors based on shared interests) are also possible (but left for future work). Fig. \ref{fig:entire_pipeline} depicts our pipeline.
(iii)~Finally, we evaluate the article recommendations produced based on our questionnaires using historical edit-history data, also investigating the cohesion and readability of questions. We then evaluate our method's article recommendation performance online, with actual Wikipedia newcomers as participants. Our results show that our method beats recommendation baselines based on article popularity, a commonly used signal. Its performance comes relatively close, especially in terms of reducing undesirable recommendations, to a collaborative filtering ceiling---which is in fact unsuitable for a real cold-start scenario since it requires user histories.

The remainder of this paper is organized as follows: We first discuss related work on recommender systems in \Secref{sec:Related work}, particularly recommender systems for Wikipedia. Afterwards, we discuss our questionnaire generation approach and article recommendation method, followed by our three approaches to extracting topics from Wikipedia articles, in \Secref{sec:Proposed method}. We then present the results of our offline evaluation in \Secref{sec:Offline evaluation}, followed by our online evaluation in \Secref{sec:Online evaluation}, and we discuss our results from both evaluations. At last, we discuss some of the nuances of deploying our system in practice, describe other uses of our questionnaire, and point towards directions for future work in \Secref{sec:Discussion}.

\section{Related work}
\label{sec:Related work}

\subsection{Types of feedback in recommender systems}
Generally speaking, two types of feedback exist in recommender systems: explicit, and implicit \cite{c1}. Explicit feedback refers to ratings that users can provide to indicate that they like or dislike an item. Implicit feedback refers to interactions that do not involve explicit statement of like or dislike, \emph{e.g.}, browsing history, purchase history, number of listens for a particular song, etc. This means that like/dislike information has to be inferred from the available implicit data, usually less reliably than explicit ratings, since the line between a user disliking an item, and that user simply not knowing about the item, becomes blurred. Wikipedia lacks explicit ratings from editors on the articles they like or dislike, and as such, only has implicit feedback.

\subsection{Recommender systems for Wikipedia}
There exist several works proposing recommender systems tailored for Wikipedia, which aim to expand Wikipedia in different ways. SuggestBot \cite{suggestbot} matches people with editing tasks in a personalized fashion based on their editing history. \textcite{wulczyn2016} expand Wikipedia by recommending articles that are present in one language and missing in another, while \textcite{piccardi2018} recommend new sections to be added to articles. However, none of these systems address the cold-start problem for editors; \textcite{piccardi2018} focus on recommending the correct sections for a given article, while \textcite{suggestbot} and \textcite{wulczyn2016} require a user's editing history to recommend articles to them, hence being unsuitable for a complete newcomer to Wikipedia.
\subsection{Questionnaire methods}
Questionnaire methods for the user cold-start problem ask the user a few questions and create an initial profile for them, so that the system can recommend items to the user right from the start. In addition to the relevance of those items, it is important for the questionnaire to be short, to avoid tiring the user \cite{c7}.

To our knowledge, the existing methods all work on explicit feedback systems - thus making them much less applicable to implicit feedback settings like Wikipedia - and each of their questions is concerned with the user's like or dislike of (or sometimes, lack of knowledge about) a single item. These methods can broadly be divided into two categories: non-adaptive, and adaptive. In adaptive methods as opposed to the non-adaptive ones, the question posed to the user at each step depends on their answers to the previous ones.

\textbf{Non-adaptive methods} usually use an information theoretic measure on all the items to choose the top n items, which will then presented to the users as questions of the type ``Do you know this item and do you like it?". Some of these measures \cite{c7,c8} are as follows:
\begin{itemize}
\item Popularity: Items are chosen based on their popularities (\emph{i.e.} how many ratings they have), a popular item being preferred over more obscure ones.
\item Contention (or Entropy): Items are chosen based on how much users disagree about them. The use of entropy without popularity may result in esoteric and obscure items being selected \cite{c9}; thus, methods such as Entropy0 (that takes missing values into account as zeros) or HELF (Harmonic mean of Entropy and Logarithm of Frequency) have superior performance to raw entropy \cite{c7}. Unfortunately, the lack of reliable negative ratings makes entropy-based methods become much less applicable in an implicit feedback setting.
\item Coverage: As noted by \textcite{c8}, certain items might be controversial, but their ratings might be weakly correlated with those of others, and therefore the rating the user gives them will not be indicative of what ratings they would probably give to other items.
\end{itemize}
\textbf{Adaptive methods} are usually based on decision trees \cite{c7,c8,zhou2011}, while some are based on bootstrapping using item-item similarity \cite{c9,c13}. Some methods fail to improve accuracy over the best non-adaptive methods (popularity-times-entropy in case of \cite{c13}), while others \cite{c7} beat methods such as HELF and Entropy0 in an online setting. \textcite{sun2013} use a decision tree with multiple questions at each step, combining the user's answers to those questions for the selection of the next node. Their multi-question approach is designed to increase the probability that the user would know at least one of the items in question in each node.

The performance of these methods can be evaluated both offline and online. Offline evaluation is the most common type, wherein the model is trained on a training set, and then the questionnaire-answering process is simulated with users in the test set. Each data point is a user's rating for an item and the training and test sets have no users in common. The test error is the mean of the prediction errors for individual user-item-rating tuples.
In online testing as used by \textcite{c7}, actual users answer the questionnaire, and their answers are treated as their profile; their subsequent ratings of recommended movies are treated as test data.

\section{Proposed method}
\label{sec:Proposed method}

\begin{figure}[t]
    \resizebox{\columnwidth}{!}{
        \includegraphics{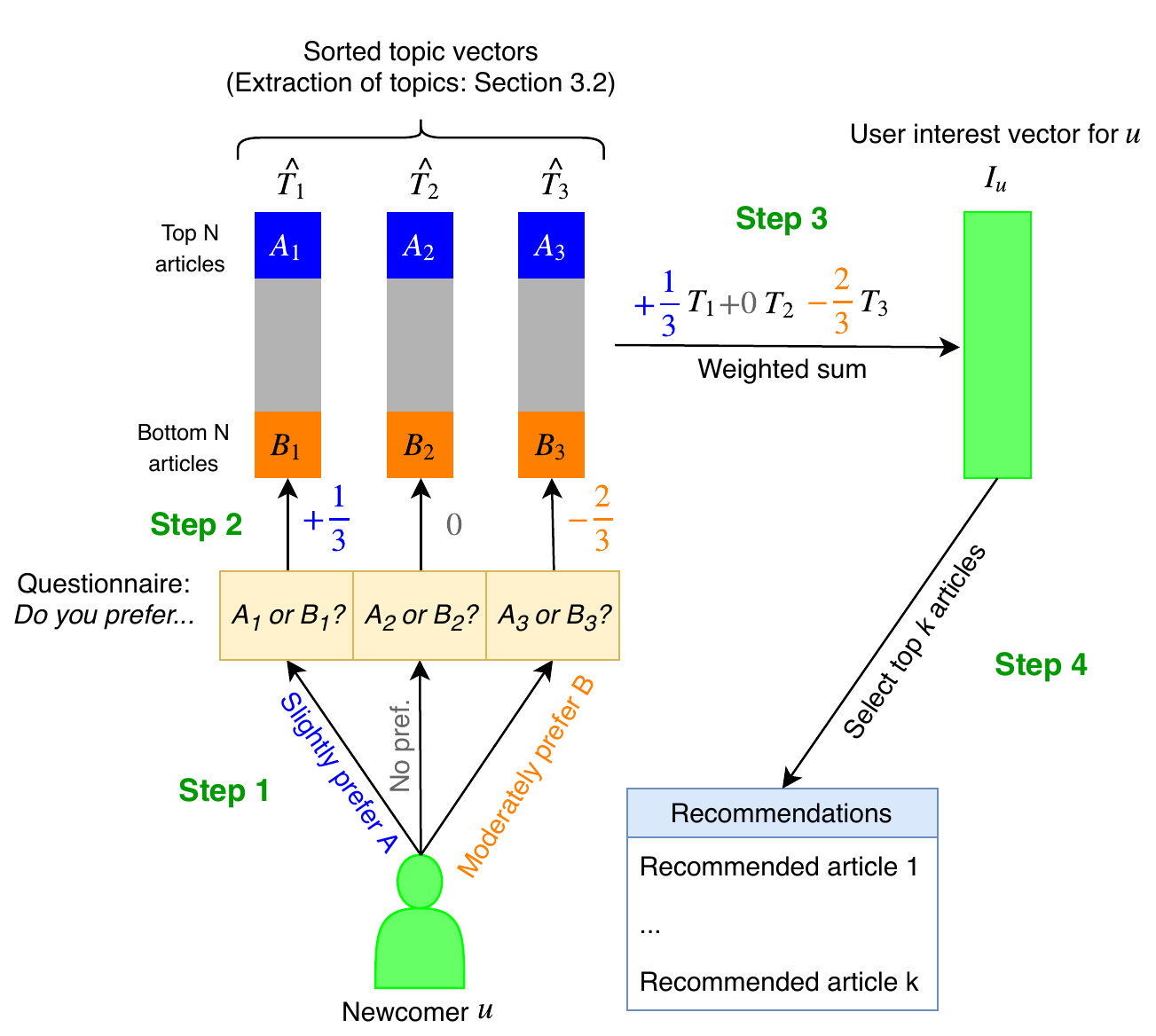}
    }
    \caption{Overview of the process of a user answering the questionnaire and the generation of recommendations using those responses. The steps mentioned in the figure are those explained in \Secref{sec:Generating recommendations from questionnaire answers}. The vectors $\hat{T_i}$ are the original topic vectors $T_i$ with their entries sorted descending from top to bottom. Both the topic and interest vectors have one entry per Wikipedia article, capturing the relevance of each article to the topic or the user, respectively.
    }
    \label{fig:method_flowchart}
\end{figure}

Our method works by first eliciting a new user's interests through a short questionnaire and then recommending articles that are well aligned with the inferred interests.

The questionnaire is generated ahead of time in a preprocessing step.
It consists of $K$ questions, each of which asks for the user's preference among two lists of articles, A and B: \textit{``Between lists A and B, which one contains more articles that you would be interested in editing?''}
Answers are collected on a 7-item Likert scale, indicating a great, moderate, or slight preference for one of the lists or a lack of preference.

For generating the questions, we require a set of \emph{topic vectors} $T_i$.
Topic vectors must have an entry for each Wikipedia article, specifying the relevance of the respective article to the topic.
Also, topic vectors must capture topical \emph{dichotomies,} such that the articles corresponding to the largest positive entries form a cluster that is clearly distinct from another cluster formed by the articles corresponding to the largest negative entries (\eg, singers \vs{} beetles, \cf example of \Tabref{tab:questions}).
For instance, the topics discovered by a singular value decomposition (SVD) on a user--article interaction matrix meet this requirement, since they define topics as directions of maximum variance among users (represented as vectors with entries corresponding to articles).
In this setup, each topic vector $T_i$ can be turned into a question $Q_i$
by simply using $T_i$'s highest\hyp scoring $N$ articles as list A and its lowest\hyp scoring $N$ articles, as list B.%
\footnote{
We choose $N=20$, which gives a good trade-off between quick readability and
sufficient context (\cf evaluation of \Secref{sec:Question cohesion}).
}

Any method producing such dichotomous topic vectors can be plugged into the above procedure as a black box.
We outline three particular methods later (\Secref{sec:Preprocessing: Extracting topic vectors}); for now, we assume the $K$ questions are given and focus on how the user's answers to the questions are leveraged for recommending articles to edit (\Secref{sec:Generating recommendations from questionnaire answers}).

\subsection{Generating recommendations from questionnaire answers}
\label{sec:Generating recommendations from questionnaire answers}

The procedure for generating recommendations from questionnaire answers consists of four steps (depicted schematically in \Figref{fig:method_flowchart}):

\begin{enumerate}
    \item User $u$ answers the $K$ questions (all of the form, \textit{``Between lists A and B, which one contains more articles that you would be interested in editing?''}) on a 7-item Likert scale.
    \item The categorical answers are turned into numerical scores between $-1$ and $1$ (great, moderate, and slight preference for A are mapped to $1$, $\frac{2}{3}$, $\frac{1}{3}$, respectively; negative values are used if the preference is for B, and $0$ is for no preference).
    \item The numerical scores serve as weights for combining the topic vectors $T_1,\dots,T_K$ in a weighted sum, which we call the \emph{interest vector} $I_u$ of user $u$. Formally, the interest vector is calculated as
    \begin{equation}
    I_u = \sum_{i=1}^{K} \rho_{ui} T_i,
    \end{equation}
    where $\rho_{ui}$ is the numerical representation (\cf\ item~2 above) of $u$'s response to the $i$-th question.
    \item Recommend to $u$ the articles corresponding to the largest entries of the interest vector $I_u$.
\end{enumerate}

This procedure is very similar to collaborative filtering methods based on matrix factorization (\eg, SVD-based methods):
For a given user $u$, those methods consider a vector $I_u^*$ with an entry for each article, storing how much $u$ has previously interacted with (\eg, edited) the respective article.
This vector is projected into a low\hyp dimensional subspace spanned by a small number of ``topic'' vectors (via the dot product of $I_u^*$ with each topic vector, capturing how well $u$ is aligned with each topic) and then back into the original space (via a weighted sum of topic vectors, where weights correspond to the previously computed dot products).
Projecting the vector $I_u^*$ back and forth smoothes the vector, making values that were previously equal to zero, non-zero, and thus giving all articles (even those the user has not interacted with) a meaningful score.

Our procedure can be seen as an approximation of this approach:
Since we deal with newcomers, $I_u^*$ is not available, so we cannot compute the dot product of $I_u^*$ with the topic vectors.
Instead, we ask $u$ explicitly how well she is aligned with each topic vector and use her responses (interpreted as scores between $-1$ and $1$) as proxies for the dot products.
From here on, our method is identical to the above\hyp described matrix\hyp factorization approach, so our resulting user interest vectors $I_u$ may be considered approximations of the smoothed vectors that would be output by that approach if we had access to historical interaction vectors $I_u^*$.

\subsection{Preprocessing: Extracting topic vectors}
\label{sec:Preprocessing: Extracting topic vectors}

As stated at the beginning of \Secref{sec:Proposed method}, our questionnaire generation method relies on topic vectors containing an entry for each Wikipedia article where the sets of highest- and lowest\hyp scoring articles form dichotomous clusters.
In this subsection, we will describe three approaches for extracting such topic vectors, which we term
``content-only'', ``collaborative-only'', and ``joint''.
Preprocessing steps for preparing the data may be found in the appendix.

\subsubsection{Content-only} Our content-only approach uses the textual content of Wikipedia articles as data, obtained from the English Wikipedia dump.%
\footnote{https://dumps.wikimedia.org/}
We use a bag-of-words representation of each article with TF-IDF weighting to create a matrix $T$ whose columns are articles and whose rows are unigrams.\footnote{Word embedding approaches such as word2vec \cite{word2vec} may also be used in lieu of TF-IDF.}
Based on the singular value decomposition (SVD) $T = USV^{T}$ (where $U$ and $V$ are orthogonal matrices and the diagonal $S$ contains the singular values in sorted order), we define the topic matrix as
$\bar{T} = VS^{T}$.%
\footnote{
Since $T$ is large and sparse, we perform the SVD without first mean-centering it.
As a consequence, the
singular vectors (columns of $U$ and $V$)
corresponding to the largest singular value capture mostly the mean of the
data
and we hence discard the first
column (\ie, topic vector) of
$\bar{T} = VS^{T}$.
}
We may now use the first $K$ columns of $\bar{T}$ to create $K$ questions using the question generation method outlined in the beginning of \Secref{sec:Proposed method}. This method essentially captures directions of high variance in terms of the usage of words across articles, and since it outputs orthogonal directions in space, the correlation between the topics is minimized.

The disadvantage of only using the content is that many words may have multiple meanings (\emph{e.g.} proper nouns that refer to different entities, such as the name ``Michael''), which could create similarities between dissimilar articles. This could result in the reduction of readability and semantic cohesion in the questions.

\subsubsection{Collaborative-only} The full editing history of Wikipedia%
\footnote{https://www.mediawiki.org/wiki/API:Revisions}
provides rich insights into how editors contribute to the project.
Based on this dataset, we create a matrix $E$ whose rows are users and whose columns are articles, with $E_{ij}$ being the number of edits user $i$ has made to article $j$.

Our collaborative-only topic extraction method is similar to our content-only method, except that the SVD is performed on the editing history matrix $E$. This approach captures directions of high variance in terms of the editors contributing to articles. Our hope is that there will exist editor communities, each of which focuses on a specific set of topics, thus allowing us to capture those topics by finding these high-variance directions. The disadvantage of this method, however, is that there will still be many users whose contributions are not necessarily semantically coherent (since many edits may be about fixing typos, \etc \cite{suggestbot}). This might cause semantically incoherent articles to be grouped together in one topic. In addition, edit counts alone may not be sufficiently representative of an article's importance, because many important articles will have fewer edits because they lack controversy, are not targeted by vandalism, or require more domain expertise to edit.

\subsubsection{Joint topic extraction method}
In order to leverage the advantages of both types of data at our disposal, we propose a joint method that embeds both users and articles in a latent space using matrix factorization. We call the matrix containing the article latent representations (in a row-wise fashion) $Q$ and the matrix with the user latent representations $P$. Our loss function is as follows:
\begin{equation} \label{objectivef1}
\begin{split}
\min_{P,Q} \; \sum_{u,i} c_{ui}(r_{ui} - p_{u}^{T}q_{i})^{2} &+ \alpha \|P\|^{2} + \lambda \|Q-\bar{Q}\|^{2} \\
&+ \theta \|Q^{T}Q - \operatorname{diag}(Q^{T}Q)\|^{2},
\end{split}
\end{equation}
where $\alpha$, $\lambda$, and $\theta$ are hyperparameters set by a grid search using a validation set (which will be explained later). The matrix $\bar{Q}$ is obtained by stacking the first 200 columns of $\bar{T}$ (\ie, content-based topic vectors) and row-normalizing the result, and serves as a regularizer for our $Q$ matrix; it is our way of including the content of articles in the objective function. The reason for the row-normalization will be explained in the optimization section. The last term in the function is designed to enforce an orthogonality constraint on the columns of $Q$ in a soft manner; this is to minimize correlation between dimensions.
The $r_{ui}$ and $c_{ui}$ are derived from the respective entry of the user-article editing history matrix, $e_{ui}$, following \textcite{c1} as
\begin{eqnarray}
    r_{ui} &=& 
\begin{cases}
    1   & \text{if } e_{ui}>0, \\
    0   & \text{otherwise,}
\end{cases} \\
c_{ui} &=& 1 + \kappa \, \ln\left(1+\frac{e_{ui}}{\epsilon}\right).
\end{eqnarray}
where $r_{ui}$ is the inferred rating by user $u$ for article $i$, while $c_{ui}$ indicates how confident we are in the inferred rating. The hyperparameters $\kappa$ and $\epsilon$ determine how quickly confidence grows with increasing edit counts; we have set them to 10 and 20 respectively as we believe these to be reasonable values, reducing the number of hyperparameters for our grid search.

As mentioned before, the number of latent dimensions for $P$ and $Q$ is 200. As with other matrix factorization problems, this objective function is non-convex. It is important to mention that the existence of the $c_{ui}$ and $r_{ui}$ means that our joint method does not reduce to the collaborative-only method if $\bar{Q}$ is set to 0.
\subsubsection{Optimization and evaluation metrics}
We optimize the objective function following
\textcite{c14}
and replacing stochastic gradient descent with mini-batch gradient descent. Since our primary aim is to extract topics, our focus is more on our more reliable information, \emph{i.e.}, the non-zero entries. Therefore, at each iteration, we take a mini-batch from our $E$ matrix that is $90\%$ non-zeros and $10\%$ zeros.

Our objective function being non-convex means that we are likely to arrive at a local - rather than global - minimum using gradient descent, and our initialization step will be an important one. We set $Q$ to $\bar{Q}$ and randomly initialize $P$, and then we row-normalize $P$, such that all three matrices are row-normalized. This is because the $r_{ui}$ in our objective function are either $0$ or $1$, and if the rows of $P$ and $Q$ are normalized, the value of the dot product between a row of $P$ and a row of $Q$ ranges between $-1$ and $1$. This does mean that the columns of $\bar{Q}$ are not orthogonal anymore, but our results show that the orthogonality constraint on $Q$ still results in almost-orthogonal columns for $Q$, and our observed results are superior to the case where we do not use row-normalization.

The values of our hyperparameters are determined by taking a validation set out of our user-article matrix. The validation set is created by choosing a fraction of the users and then taking out a fraction of their edits, and then training the objective on the remaining user-article matrix. In the end, we calculate two values for the selection of hyperparameters:
\begin{itemize}
\item The validation error, which is the mean squared error (\emph{i.e.} the first term in Eq. \ref{objectivef1}) on the held-out edits for the validation set users.
\item The cohesion score for $Q$, that is calculated as follows: for each topic, we calculate the cosine between each pair of article latent vectors for the set of the 20 most positively weighted articles in that topic and similarly for the set of the 20 most negatively weighted articles, and we then take the average separately for each of the two sets. The average of these two values is the cohesion score for that topic, which has a maximum of 1. This score is an important measure because the questions we generate and the dichotomies they capture need to be easily intelligible to the questionnaire respondent. We will also use this cohesion score to compare the quality of the questions generated by our three methods.
\end{itemize}
To choose the best hyperparameters, we look at these (error,score) pairs and identify the point that offers the best trade-off between a low validation error and a high cohesion score.

\begin{table}[t]
\caption{The mean and 95\% confidence interval of cohesion for 20 and 50 questions generated using our three methods.}
\label{tab:offline_cohesion_table}
\resizebox{\columnwidth}{!}{%
\begin{tabular}{|l|l|l|}
\hline
Method              & Avg. cohesion, 20 qs & Avg. cohesion, 50 qs \\ \hline
Joint & $0.672 [0.626,0.719]$       & $0.674 [0.627,0.719]$       \\ \hline
Collab-only & $0.578 [0.494,0.665]$       & $0.582 [0.533,0.632]$       \\ \hline
Content-only & $0.434 [0.375,0.493]$       & $0.439 [0.388,0.493]$       \\ \hline
\end{tabular}%
}
\end{table}

\begin{table}[t]
\caption{The mean and five precentiles of recall at 300 for the three baselines and our three methods (content-only, collaborative-only, and joint), with the addition of a collaborative-only method without the septile-based stratification (Collab-only, no strat.) in the question answering simulation. Our methods are in bold. The total number of test users is 15,028.}
\label{tab:offline_simulation}
\resizebox{\columnwidth}{!}{%
\begin{tabular}{|l|r|r|r|r|r|r|r|}
\hline
Method &      mean &   50\% &   75\% &   90\% &   95\% &   99\% &  99.9\% \\ \hline
CF-based &  0.06758 &  0.05 &  0.10 &  0.20 &  0.25 &  0.45 &   0.70 \\ \hline
{\bf Content-only} &  0.02113 &  0.00 &  0.05 &  0.05 &  0.10 &  0.20 &   0.40 \\ \hline
Edit-pop &  0.01081 &  0.00 &  0.00 &  0.05 &  0.05 &  0.10 &   0.20 \\ \hline
{\bf Joint} &  0.01077 &  0.00 &  0.00 &  0.05 &  0.05 &  0.15 &   0.30 \\ \hline
{\bf Collab-only, no strat.} &  0.00681 &  0.00 &  0.00 &  0.05 &  0.05 &  0.10 &   0.25 \\ \hline
View-pop &  0.00490 &  0.00 &  0.00 &  0.00 &  0.05 &  0.05 &   0.10 \\ \hline
{\bf Collab-only} &  0.00488 &  0.00 &  0.00 &  0.00 &  0.05 &  0.10 &   0.20 \\ \hline
\end{tabular}
}
\end{table}

\begin{table}[t]
\caption{Top and bottom 5 articles in the first 3 questions, generated by the joint method.}
\label{tab:questions}
\resizebox{\columnwidth}{!}{%
\begin{tabular}{l|l|l|l|}
\cline{2-4}
                               & Question 1                                                                                                            & Question 2                                                                                                                                            & Question 3                                                                                                                                                 \\ \hline
\multicolumn{1}{|l|}{Top 5}    & \begin{tabular}[c]{@{}l@{}}Meat Loaf\\ Elvis Costello\\ Stevie Nicks\\ Eddie Vedder\\ Tom Jones (singer)\end{tabular} & \begin{tabular}[c]{@{}l@{}}Sikeston, Missouri\\ Selma, Alabama\\ Timeline of Baltimore\\ St. Augustine, Florida\\ Clarksville, Tennessee\end{tabular} & \begin{tabular}[c]{@{}l@{}}Philipp, Prince of Eulenburg\\ Maximilien Robespierre\\ Charles George Gordon\\ In Solitary Witness\\ Mortara case\end{tabular} \\ \hline
\multicolumn{1}{|l|}{Bottom 5} & \begin{tabular}[c]{@{}l@{}}Apomecynini\\ Desmiphorini\\ Agaristinae\\ Endemic goitre\\ C21H20O10\end{tabular}         & \begin{tabular}[c]{@{}l@{}}Incomplete Nature\\ Troubleshooting\\ Affective computing\\ Isothermal microcalorimetry\\ Positive feedback\end{tabular}   & \begin{tabular}[c]{@{}l@{}}Tri-Cities, Washington\\ Kent, Ohio\\ Champaign, Illinois\\ Limon, Colorado\\ Carlsbad, New Mexico\end{tabular}                 \\ \hline
\end{tabular}
}
\end{table}

\section{Offline evaluation}
\label{sec:Offline evaluation}

\subsection{Question cohesion}
\label{sec:Question cohesion}

Our first approach to evaluating our questions is calculating the cohesion of the questions. Table \ref{tab:offline_cohesion_table} shows the cohesion values for 20 and 50 questions for the joint method, the content-only SVD, and the collaborative-only SVD. The cohesion scores here, in each of the three schemes, have been calculated using the same matrix for both the questions and the latent representations, thus in some sense simultaneously measuring both the quality of individual questions and their latent representations. We also manually check the first 20 questions generated by the method. Our manual investigation reveals that the content-only method has a few relatively outlier articles in each 20-article list, which could confuse the user regarding the meaning of the question. The top and bottom 5 articles for the first 3 questions can be seen in \Tabref{tab:questions}.

\subsection{Recommendation performance} The second element of our offline evaluation is recommendation performance; we perform it by simulating users in the test set answering the questionnaire. This simulation is as follows: for each test user, we hold out 20 of their distinct edited articles, \emph{i.e.} in their edit vector, we set to zero 20 of the (non-zero) entries, creating a ``modified edit vector". Then, in order to simulate the online experiment, we calculate the dot product between each question vector and the user's modified edit vector - which results in a quantity we call the user's simulated response to that question - and then calculate the 1st to 6th septiles of the set of all these values across all users. These septiles will allow us to rank-normalize the users' simulated responses to the questions into the 7-level Likert scale that we discussed earlier. This gives us a response vector for the user, consisting of all their simulated responses. This method is equivalent to projecting the user's article-space edit vector onto the subspace spanned by the topics (\emph{i.e.} questions), and then representing the resulting vector back in the original basis, with the difference that we also discretize their projection onto each dimension of the subspace.

We generate the user's recommendations as described previously. Using this list, we calculate recall@$k$ (the fraction of the holdout article set appearing in the top $k$ recommendations) and precision@$k$ (the fraction of the top $k$ recommendations appearing in the holdout article set) for the 20 holdout articles of the user in question. Note that since the total number of true positives is known (there are 20 holdout articles), recall@$k$ and precision@$k$ are trivially related by a known fraction and therefore, we only report recall@$k$.

Due to the sparsity of the modified edit vector, our simulation of the question answering is using the full topic vector, rather than a topic vector with everything except the top and bottom 20 articles masked; such a masking would have resulted in most responses being 0. This, combined with our method for evaluation (the 20-article holdout set), means that the offline evaluation is relatively unrealistic.

We use two baselines and one ceiling to compare the performance of our method with:
\begin{itemize}
\item Recommendation based on editing popularity, where we recommend the most edited articles. This is a weak baseline since it is not personalized for each user. We call this baseline ``Edit-pop''.
\item Recommendation based on viewing popularity, where we recommend the most viewed articles. This is also a weak baseline. We call this baseline ``View-pop''.
\item Recommendation using collaborative filtering (with 50 latent dimensions, selected using a validation set) on the user-article editing matrix, which we call ``CF-based''. This is done using the Python package \emph{implicit}\footnote{https://github.com/benfred/implicit}, using an objective function proposed in \cite{c1}, which is essentially our own with $\bar{Q}$ set to 0 and the soft orthogonality constraint removed.
\end{itemize}

The collaborative filtering (CF) method is a ceiling and not a baseline, because the primary target audience of a cold-start questionnaire comprises users who have no editing history since they have just created an account, and CF is unusable in the absence of existing history for the user. As such, a comparison of our method's performance with the performance of CF for users for whom there is some existing editing history, allows us to gauge how far we are from a potentially unattainable high level of performance.

The results can be seen in Table \ref{tab:offline_simulation}. We have additionally included the collaborative-only method without the septile-based stratification (whose absence would make the offline evaluation less realistic) to show the negative effect of said stratification on offline recommendation performance. Interestingly, the content-only method outperforms our joint method in the offline simulation, and is only outperformed by the CF-based recommendations. This is especially surprising because the holdout set of users is essentially part of the greater editing history matrix, which plays no role in the content-only method, but does play a role in our joint method. It is also worth noting that the collaborative-only method is the worst-performing method. This, combined with the cohesion values we see in Table \ref{tab:offline_cohesion_table}, means that the editing history matrix itself is noisy and reduces our offline performance in the joint method; however, it also helps denoise the questions generated by our method, making them much more cohesive. This could be because dissimilar articles that possess similar terms do not have similar editor communities, and this allows our joint method to separate these dissimilar articles from each other (while the content-only approach would fail to do so).

As mentioned before, our offline test is rather unrealistic because of how the question answering process is simulated and how the results are evaluated, even though it is a necessary sanity check for the output of our method. An online evaluation with real users will be a more realistic evaluation of the system, and we will discuss how we have conducted this online evaluation in the next section.

\section{Online evaluation}
\label{sec:Online evaluation}

Our online evaluation will attempt to measure the satisfaction of recently joined users with the recommendations provided by our method, in comparison with two baselines and a ceiling. For this evaluation, we will have to choose one of the three question sets, based on the offline evaluation. Our joint method provides the best trade-off between question cohesion and offline recommendation performance, given that it performs acceptably in the latter while being by far the best in the former. As a result, we will use the questions generated using topics extracted by our joint method in our online evaluation.

Compliant with Wikimedia's ``Wikipedia is not a laboratory'' policy, our research project has been fully documented on Wikimedia's meta-wiki\footnote{\url{https://meta.wikimedia.org/wiki/Research:Voice_and_exit_in_a_voluntary_work_environment/Elicit_new_editor_interests}} as a research project, and we have taken cautionary steps in order not to disrupt the normal operation of any Wiki in any shape or form.
\subsection{Design}
We took several steps before launching the experiment itself, detailed below:
\begin{itemize}
    \item We took all the users who had signed up for an account on the English Wikipedia in the 3 months from September to December 2018,%
    \footnote{A link to the privacy policy for the experiment is found on the meta page, under ``Second Online Experiment''.}
    since we wanted our participants to be relatively new to the platform, and not seasoned veterans. We divided these users into two sets: those with more than 3 edits, and those with 3 or fewer edits. For the former set, we have three baselines: View-pop, Edit-pop, and CF-based, while for the latter, we only have View-pop and Edit-pop. We will call the former condition ``with CF'', while the latter will be called ``no CF''.
    \item In order to calculate the CF-based recommendations for the first set of users, we first filtered their edits to only include \emph{articles} (rather than talk/user pages). Then, similarly to what we did for our offline evaluation, we used our user-article editing history matrix with the Python package ``implicit'' (again with 50 latent dimensions) to pre-calculate the recommendations, which we then stored. At this point, the ``with CF'' condition numbered about 26,000 users. For each of these users, we calculated a unique participation token that would allow us to identify them when they took the questionnaire, so that we could give them their pre-calculated CF recommendations when they participated.
    \item We subsampled the set of ``no CF'' users to also contain 26,000 users, and also generated participation tokens for them, for the sake of homogeneity.
\end{itemize}

\begin{table}[t]
\caption{Number of ``wins'', ``draws'' and ``losses'' for our method versus the three competitors for the reading preference question in each of the conditions.}
\label{tab:reading_pref_ternarised}
\resizebox{\columnwidth}{!}{%
\begin{tabular}{|l|c|c|c|c|c|c|}
\hline
\multirow{2}{*}{Reading preference} & \multicolumn{3}{c|}{with CF} & \multicolumn{3}{c|}{no CF} \\ \cline{2-7} 
                                    & losses    & draws   & wins   & losses   & draws   & wins  \\ \hline
Q-based vs CF-based                 & 172       & 73      & 115    & -        & -       & -     \\ \hline
Q-based vs View-pop                 & 108       & 70      & 182    & 81       & 53      & 163   \\ \hline
Q-based vs Edit-pop                 & 55        & 94      & 211    & 53       & 65      & 179   \\ \hline
\end{tabular}%
}
\end{table}

\begin{table}[t]
\caption{Number of ``wins'', ``draws'' and ``losses'' for our method versus the three competitors for the editing preference question in each of the conditions.}
\label{tab:editing_pref_ternarised}
\resizebox{\columnwidth}{!}{%
\begin{tabular}{|l|c|c|c|c|c|c|}
\hline
\multirow{2}{*}{Editing preference} & \multicolumn{3}{c|}{with CF} & \multicolumn{3}{c|}{no CF} \\ \cline{2-7} 
                                    & losses    & draws   & wins   & losses   & draws   & wins  \\ \hline
Q-based vs CF-based                 & 178       & 79      & 103     & -        & -       & -     \\ \hline
Q-based vs View-pop                 & 103       & 88      & 169    & 68        & 74      & 155   \\ \hline
Q-based vs Edit-pop                 & 46        & 113     & 201    & 50        & 92      & 155   \\ \hline
\end{tabular}%
}
\end{table}

At this point, the experiment was ready to begin, and we set it up as follows:
\begin{itemize}
    \item We sent emails to users from both sets, containing an invitation message, a link to the questionnaire, and their participation tokens. We attempted sending emails to all the users in both sets, but some had not enabled the receipt of emails, and only 13,794 users in total could be contacted.
    \item Each user answered the questionnaire, which consisted of the first 20 questions generated by the joint topic extraction method, where each question was about comparing two lists of 20 articles each and indicating the list that contained more articles that they would be interested in editing. For each list of articles, a word cloud of the most important terms in that list of articles was provided to provide further summary information about that list of articles, in addition to the articles themselves. Without the word clouds, some article lists (from different questions) would appear to be about the exact same subject, whereas the word clouds would reveal nuanced differences between them. For each question, the user was given the seven options as described earlier on. Their responses to the questionnaire were used to construct their article-space profile and thus to generate their questionnaire-based recommendations.
    \item For each user, twelve lists of recommendations were generated, each of which contained 5 articles. For all the users, six of these lists were generated using their responses to the questionnaire (which we will call ``Q-based'' from now on). For ``with CF'' users, two of the remaining six were View-pop, two Edit-pop, and two CF-based. For ``no CF'' users, three of the remaining six were View-pop and the other three Edit-pop. The six lists of Q-based recommendations were each paired with one of the baseline lists, resulting in six pairs of lists, which the user would then provide feedback on.
    
    Since over-specialization is a well-known issue in recommender systems \cite{mcnee2006_2} and in order to increase the diversity of our recommendations, we have used a simple diversification scheme for each of the four recommendation methods: we first generated a larger set of recommendations (randomly from the most popular articles in case of View-pop and Edit-pop), which we then clustered into several groups based on the latent representations we had generated for the articles (the rows of the $Q$ matrix), and then we randomly took one article from each cluster. This scheme was mainly meant for the Q-based and CF-based recommendations, but for the sake of a fair comparison, we also used it for View-pop and Edit-pop.
    \item For each of the six pairs of lists, the users were asked three questions (where each, again, had 7 options):
    \begin{enumerate}
        \item Among the two lists, which list has more articles that you would be interested in reading?
        \item Among the two lists, which list has more articles that you would be interested in editing?
        \item Among the two lists, which list has more articles that you would not be interested in at all, neither for reading nor for editing?
    \end{enumerate}
    \item The user's responses and their username (which was uniquely identifiable based on their participation token) were recorded.
\end{itemize}

\begin{table}[t]
\caption{Number of ``wins'', ``draws'' and ``losses'' for our method versus the three competitors for the uninterestedness question in each of the conditions.}
\label{tab:uninterestedness_ternarised}
\resizebox{\columnwidth}{!}{%
\begin{tabular}{|l|c|c|c|c|c|c|}
\hline
\multirow{2}{*}{Uninterestedness} & \multicolumn{3}{c|}{with CF} & \multicolumn{3}{c|}{no CF} \\ \cline{2-7} 
                                  & losses    & draws   & wins   & losses   & draws   & wins  \\ \hline
Q-based vs CF-based               & 137      & 114      & 109    & -        & -       & -     \\ \hline
Q-based vs View-pop               & 90      & 103      & 167     & 67      & 83      & 147    \\ \hline
Q-based vs Edit-pop               & 54      & 111      & 195     & 54      & 89      & 154    \\ \hline
\end{tabular}%
}
\end{table}

\begin{figure*}[!t]
\resizebox{\textwidth}{!}{%
\subfloat[]{\includegraphics[height=2in]{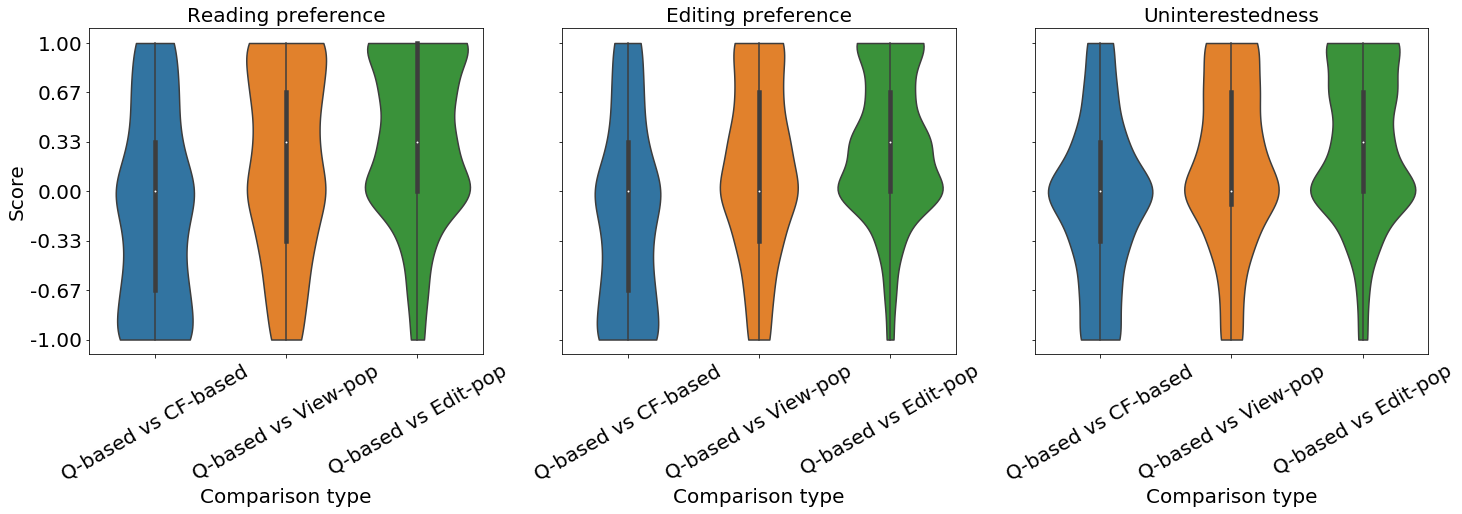}}
\subfloat[]{\includegraphics[height=2in]{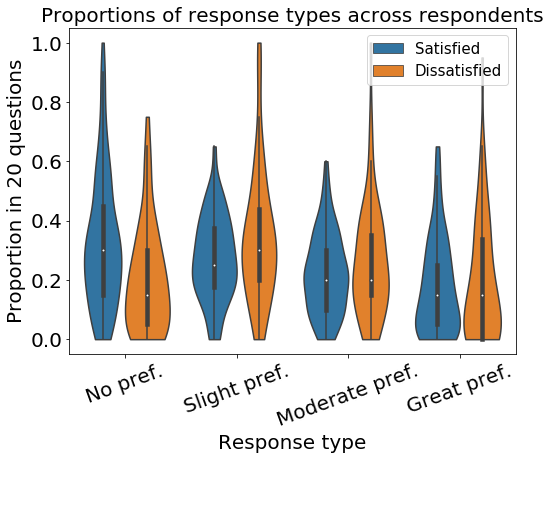}}
}
\caption{Violin plots for original responses to the three feedback questions and response type proportions to questionnaire questions.
\textbf{(a)}~Violin plots for original responses to the three feedback questions for users in the ``with CF'' condition.
\textbf{(b)}~Violin plot of questionnaire response type proportions for each response type, for strongly dissatisfied and satisfied users. The distribution of ``no preference'' responses is different between the two sets with $p < 0.001$.
(Black bar: inter-quartile range; white dot: median.)
}
\label{fig:online_violinplots}
\end{figure*}

Note that we ask separate questions for editing and reading preferences, since reading an article and contributing to it are fundamentally different, but a respondent might mix them up if they are not in separate questions. In addition, we asked the question about which list had more articles they were not interested in, since previous work has shown that users may distrust recommender systems that make recommendations that are obviously bad \cite{mcnee2006_1}.

\subsection{Results}
We received a total of 279 responses, 180 of which were from the ``with CF'' set while the remaining 99 were ``no CF''. This imbalance in the response rates is not surprising, since more active new editors (those who edit more in the first few days after creating an account) are much more likely to continue participating in Wikipedia \cite{panciera2009}.

In order to present the results, we categorize the results for each question into three types of comparisons (Q-based versus each of the three competitor methods), mapping the responses to their numerical representations from $-1$ to $1$ in steps of $\frac{1}{3}$, with the sign indicating whether the Q-based recommendations beat the other method (and $0$ corresponding to a draw). In the reading and editing preference questions, a positive value indicates that our method recommended more interesting articles than the other method, while for the uninterestedness question, a positive value indicates that our method recommended fewer uninteresting articles than the other method.

The ternarized results (win, draw and loss, corresponding to positive, zero and negative responses) for all three comparison types in each of the three questions can be seen in Tables \ref{tab:reading_pref_ternarised}, \ref{tab:editing_pref_ternarised} and \ref{tab:uninterestedness_ternarised}, showing that our method beats the popularity-based baselines for both conditions and all three questions, while the CF-based ceiling remains the best-performing method. For the ``with CF'' users, violin plots for the original response values (from -1 to 1) can be seen in Fig. \ref{fig:online_violinplots}a. Interestingly, both Table \ref{tab:uninterestedness_ternarised} and Fig. \ref{fig:online_violinplots}a show that the margin of defeat for our method against CF-based recommendations is considerably smaller in the uninterestedness question, which points towards the gap between our method and CF-based recommendations being smaller than the gap between the popularity-based baselines and our method. In particular, our method seems to be close to the CF-based ceiling when it comes to not providing undesirable recommendations, as evidenced by the results of the uninterestedness question.

\subsection{Analysis of dissatisfied users} An interesting question is the following: how many of our respondents are dissatisfied with their Q-based recommendations, and if so, are there any patterns in their recommendations or in the way they answered the questionnaire? As a response to the first question, we define three sets of respondents: strongly dissatisfied, mildly dissatisfied, and satisfied. We define strongly dissatisfied users as those who have indicated the Q-based recommendations as the less interesting ones compared to popularity-based baselines in at least half of their uninterestedness questions. These users number 58 respondents (out of 279), which means about 20\% of our respondents, pointing towards the potential existence of systematic effects. Satisfied users are those who have not done so in any of the uninterestedness questions, and they make up 135 respondents, which is about 50\% of all the respondents. Slightly over two-thirds of each of the two sets have the ``with CF'' condition. We call the rest of the users ``mildly dissatisfied'', and we exclude them from this analysis. In order to better understand the characteristics of the strongly dissatisfied users compared to the satisfied users, we analyze three features in these two sets of users:
\begin{itemize}
    \item For each user, the proportion of each of the four types of response (no preference, slight preference, moderate preference, and great preference) across all of their 20 questionnaire responses.
    \item The proportion, across each of the two sets, of their preferential and non-preferential (\emph{i.e.} non-zero and zero) responses for each question in the questionnaire.
    \item The lists of recommendations given to them, investigated manually.
\end{itemize}
Our reason for investigating \emph{how} the dissatisfied users answered the questionnaire is that our questions have different numerical latent representations in the article space, and the effects and interactions of these latent representations are unclear to us. An analysis of how the dissatisfied users differed in their questionnaire answering from the satisfied users can give us useful insights into the improvements we could make to our method.

The distribution of proportions of each response type for each of the two groups are shown in Fig. \ref{fig:online_violinplots}b. We conducted a Kruskal-Wallis H-test to see if the difference between the proportions of the ``no preference'' option between the two sets of users was significant, which gave us a p-value of 0.00014, meaning that there is a significant difference between the distributions of said proportion among the two sets of users. This tells us that the satisfied users have had a greater tendency to use the ``no preference'' option (we do not have enough evidence to state the inverse, namely that using the ``no preference'' option more would lead to greater satisfaction). This could point to our method performing worse when many non-zero responses are given, especially since the article-space representations of questions can be conflicting, \emph{i.e.} a positive response to a question up-weights some articles while down-weighting others, and these down-weighted articles may themselves have been up-weighted by another question, which creates a conflict. Thus, this is a area for future improvement.

We investigated the per-set (the sets being strongly dissatisfied and satisfied) proportions of non-preferential and preferential responses to each question, but we did not see any meaningful pattern in the differences, and as such, we cannot say that the issue comes mainly from a certain part of the questionnaire (\emph{e.g.} the earlier questions or the later questions). As the final part of our investigation, we also manually looked at the recommendations given to the dissatisfied users. However, no distinct patterns could be found in their recommendations, other than occasional over-specialization, which indicates that a more advanced diversification scheme is an important direction for future work.

\section{Discussion and future work}
\label{sec:Discussion}
In this section, we present the implications of this study and future directions.

\subsubsection{Tailoring recommendations for contribution} In the evaluation of the questionnaire, we asked editors to assess lists of recommended articles for reading and editing purposes. Such lists can be improved in two ways. First, the reading recommendations for editors may be different than reading recommendations for readers as such reading recommendations can inform the editing. Such lists are currently being generated agnostic to whether the user is an editor or a reader. Second, the editing recommendations do not consider the specific needs of the articles that are being recommended for editing. For example, articles of lower quality or stubs may need to be prioritized in such recommendations using new and existing system \cite{Halfaker:2017:IQD:3125433.3125475}. One can also provide recommendations based on the type of edits an article is missing: for example typo fixing, adding citations, expanding a section, \etc Since our method essentially creates a total ordering for the article space, it can, given a list of stubs and other articles needing contribution, limit its recommendations to those.

\subsubsection{Improving the recommendations} One direction for future work is taking into account the sizes and other details of individual edits, and the discussions on talk pages of articles, which would give us richer information on users and potentially improve our editing history matrix. Another direction would be addressing the issue of lowered recommendation quality when faced with potentially conflicting responses to the questionnaire, which could be achieved through a more advanced diversification scheme that could make sure that all questions get their fair share of presence in the recommendations while retaining the ability of our method to recommend articles that bridge different topics. Also, an interesting improvement would be making the questionnaire adaptive, thus technically allowing for more topics to be included in the questionnaire, as not everyone would be getting the same 20 questions in an adaptive questionnaire. This would allow for more detailed initial profiles for users. In addition to adaptivity, one question that could be addressed in future work is finding an optimal number of questions which creates a balance between eliciting enough information and being short enough for the user to answer quickly. We estimate (based on small-scale tests) that each question in the current questionnaire would take about one minute to answer, thus bringing up the total answering time to about 20 minutes, which could drive some newcomers away. A simple solution would be to allow the user to answer as many questions as they would like, rather than a fixed number of questions.

\subsubsection{User matching} An interesting approach to helping newcomers (and an alternative to article recommendation) is matching them with other newcomers or with veteran editors to help get them started. The latter can provide the newcomer with a mentor of sorts, while the former allows the formation of more flat and non-hierarchical groups who can collaboratively contribute to Wikipedia. Our questionnaire allows us to perform such a matching, since each user can be represented as their vector of responses to the questions, and the similarity between two users can be defined as the dot product of their response vectors. We conducted an experiment to assess how such a matching would help newcomers, but our experiment was unsuccessful due to low participation. We hypothesize that the low participation was due to the complexity of the experiment design, some of which is imposed by the way Wikipedia communications between editors work today. We believe that exploring matching based on the constructed user interest profiles is a potential direction for the future, and an interesting use of our questionnaire.

\subsubsection{Measuring retention} In this work, we focused on measuring the quality of our recommendations and the immediate satisfaction of the users with them. Another important question is whether the deployment of this cold-start recommender leads to increased retention among users who may not have become active editors otherwise. There will need to be a different and longer experiment to measure retention, potentially by providing some newcomers with personalized recommendations and others with random recommendations and comparing their contributions over time.

\section{Conclusions}
We have proposed a questionnaire creation pipeline---consisting of a novel automatic question generation method, based on pairwise comparisons, and a set of three topic extraction methods---that can create questionnaires to help tackle the editor cold-start problem on Wikipedia. We showed that our proposed joint topic extraction provides the best trade-off between the semantic cohesion of questions and offline article recommendation performance. Our online evaluation shows that our method provides article recommendations that are superior to non-personalized baselines, while having relatively close performance to a collaborative filtering ceiling which is inapplicable to a real cold-start scenario due to its need for the editing history of the user. Our method's performance is achieved using a non-adaptive questionnaire, and we believe that adding adaptiveness can boost its performance even further.

\section{Acknowledgments}

We would like to thank Tiziano Piccardi and Hristo Paskov for their assistance, and the Wikipedia editor community for participating in our online evaluation and making this study possible. We would also like to thank the Wikimedia research community for their valuable feedback.

\begin{appendices}
\section{Preprocessing the data}
To improve the results of our method, we have performed several preprocessing steps on the set of articles and the set of users before training our model, including:
\begin{itemize}
\item Removing very rare (appearing in less than 50 articles) and very frequent (appearing in more than 10\% of the articles) words.
\item Removing users and articles with very low edit counts (the lower bound being 20 in both cases).
\item Removing very short stubs (under 100 words), since their content is too sparse and noisy.
\item Removing lists and list-like articles since they a) exist more for reasons of organization rather than pure content, and b) contain many rare terms (\emph{e.g.} proper nouns) due to their nature, and according to our tests, they end up disproportionately dominating the topics if not removed. We have not conducted this filtering in an exhaustive manner since it is sometimes debatable whether or not an article is a list; a more exhaustive removal of such articles is left for future work.
\end{itemize}
Our filtered article space contains approximately 2,200,000 articles, which is still a very large set. However, if the possibility for the recommendation of any and all articles is required, then the article recommendations made by our method may be used as an intermediate step, with the final recommendations (out of the full article set) being selected based on the intermediate recommendations. This selection can be based on content similarity to the intermediate recommendations, the need for contributions and its severity, various metadata about the article (\emph{e.g.} its length, number of sections, whether or not it is a list), \etc For the sake of reproducibility and easier future expansion of our work, we have made both our code and our data publicly available.%
\footnote{\url{https://github.com/RamtinYazdanian/wikipedia_coldstart_questionnaire}}
\end{appendices}



\bibliographystyle{aaai}

\end{document}